\RequirePackage[running]{lineno}

\documentclass[aps,prl,nofootinbib,superscriptaddress,letterpaper,amsmath,amssymb]{revtex4}

\usepackage{graphicx}  
\usepackage{dcolumn}   
\usepackage{bm}        
\usepackage{amssymb}   
\usepackage{feynmf}
\usepackage{slashed}
\usepackage{multirow}
\usepackage{verbatim}
\def\gsim{\lower0.5ex\hbox{$\:\buildrel >\over\sim\:$}}
\def\lsim{\lower0.5ex\hbox{$\:\buildrel <\over\sim\:$}}

\let\g=\gamma

\newcommand{\met}{\slashed{E}_{\rm T}}
\newcommand{\be}{\begin{equation}}
\newcommand{\ee}{\end{equation}}
\newcommand{\bea}{\begin{eqnarray}}
\newcommand{\eea}{\end{eqnarray}}

\newcommand{\nbox}{{\,\lower0.9pt\vbox{\hrule \hbox{\vrule height 0.2 cm
\hskip 0.2 cm \vrule height 0.2 cm}\hrule}\,}}

\def\sub#1{_{\lower.25ex\hbox{$\scriptstyle#1$}}}

\newskip\zatskip \zatskip=0pt plus0pt minus0pt
\def\matth{\mathsurround=0pt}
\def\lsim{\mathrel{\mathpalette\atversim<}}
\def\gsim{\mathrel{\mathpalette\atversim>}}
\def\sigv{\ifmmode \langle\sigma v\rangle\else $\langle\sigma v\rangle$\fi}
\newskip\zatskip \zatskip=0pt plus0pt minus0pt
\def\matth{\mathsurround=0pt}
\def\lsim{\mathrel{\mathpalette\atversim<}}
\def\gsim{\mathrel{\mathpalette\atversim>}}
\def\atversim#1#2{\lower0.7ex\vbox{\baselineskip\zatskip\lineskip\zatskip
  \lineskiplimit
  0pt\ialign{$\matth#1\hfil##\hfil$\crcr#2\crcr\sim\crcr}}}

\begin{document}

\begin{flushright}
OHSTPY-HEP-T-15-001
\end{flushright}

\thispagestyle{empty}
\vspace*{-3.5cm}

\vspace{0.5in}

\title{
\vspace{8mm}
Searching for Standard Model Adjoint Scalars with Diboson Resonance Signatures}

\begin{center}
\begin{abstract}
We explore the phenomenology of scalar fields in the adjoint representation of SM gauge groups.
We write a general set of  dimension 5 effective operators in which SM adjoint scalars couple to pairs of standard model bosons.
Using these effective operators, we explore new possible decay channels of a scalar color octet into a gluon and a Z boson/ gluon and a photon.
We recast several analyses from Run I of the LHC to find constraints on an a scalar octet decaying into these channels, and we project the discovery potential of color octets in our gluon+photon channel for the 14 TeV run of LHC.

\end{abstract}
\end{center}

\author{Linda M. Carpenter}
\affiliation{The Ohio State University, Columbus, OH}
\author{Russell Colburn}
\affiliation{The Ohio State University, Columbus, OH}

\pacs{}
\maketitle


\section{Introduction}

With Run II of the LHC now in progress, we are now able to probe well into the TeV range in the search for new particles.  One discoverable class of heavy particles which appears in many Beyond the Standard Model (BSM) scenarios is that of scalar fields in adjoint representations of the Standard Model (SM) gauge groups.
Scenarios with scalar adjoints are quite diverse and candidate fields include chiral states that mix with gauginos in R-symmetric extensions of the Minimal Supersymmetric Standard Model \cite{Fox:2002bu}\cite{Hall:1990hq}, triplet Higgs fields, various states in Minimal Flavor Violating models \cite{Manohar:2006ga}, and KK partners of gauge bosons in extra-dimensional models \cite{Randall:1999ee}.
Proposed search channels for these states vary depending on the scenario, but include such signatures as jets plus missing energy, $t\overline{t}$ and dijet resonances, and searches for W/Z + jets \cite{Plehn:2008ae}\cite{Choi:2009ue}\cite{Chen:2014haa}\cite{Carpenter:2011yj}.


In this work we will explore the consequences of couplings between scalar adjoint fields and pairs of SM gauge bosons. We will begin in the framework of Effective Operator analysis.
As we have seen from recent studies in Higgs physics and Dark Matter phenomenology, this is an extremely useful tool for systematically cataloging all possible interactions consistent with symmetries \cite{Carmi:2012in}\cite{Goodman:2010yf}.
We first present the most general effective Lagrangian up to dimension 5 in which scalar fields in adjoint representations of the SM gauge groups couple to pairs of gauge bosons. These operators are allowed by all symmetries of the theory, we will therefore assume they are generically non-zero.  Some dimension 5 operators involving scalar adjoints have been well studied, for example the dimension 5 coupling between a scalar color octet and two gluons.
Here, we will present several additional operators which induce new di-boson couplings.

We also discuss possible UV completions of the effective theory in several scenarios.  These completions will involve integrating out a messenger sector of fields which are charged under multiple SM gauge groups.
We will consider both supersymmetric and non-supersymmetric completions of the effective operators.
We will see that for typical completions, for example simple messenger sectors consisting of vector-like fermions, gauge invariance demands the simultaneous existence of multiple dimension 5 operators. In general individual operators cannot be arbitrarily ignored.  We then study the collider phenomenology induced by the operators in our effective Lagrangian.

We choose to explore the phenomenology of a scalar color octet whose production and decay proceeds through loop-level couplings to pairs of Standard Model bosons.  
In particular we focus on the effective couplings between the scalar octet and a gluon and photon/Z-boson. 
We argue that new proposed decay channels for the octet opens a window to constrain or discover various models of octet scalars.  
We recast several 8 TeV LHC analyses to find current constraints on effective operators coefficients.  
These analyses include searches for dijet resonances, searches for excited quarks, and the inclusive search in the mono-jet channel.  
We then present the discovery potential for the scalar octet in a jet plus photon channel for the 14 TeV run of LHC.

This paper proceeds as follows, in section 2 we consider effective operators where a scalar SM adjoint may couple to gauge bosons.  
In section 3 we discuss possible UV completions of the effective Lagrangian.
In Section 4 we focus on the production and decay modes of a scalar color octet at LHC given our set of effective operators.
In section 5 we present the bounds on scalar color adjoints in our scenario coming from a recast of existing searches LHC Run 1.
In section 6 we present a sensitivity search for scalar adjoints in $S \rightarrow g \gamma$ channel in the 14 TeV run of LHC.  Section 7 concludes.

\section{Effective Operators}

We will now explore the couplings of SM adjoint scalars to pairs of gauge bosons, beginning with operators of the lowest dimension.  We consider two new spin 0 adjoint fields, color octet, S, with quantum numbers $(8,1)$ under SM gauge groups, and a weak triplet T with quantum numbers $(1,3)$ under SM gauge groups.
At dimension 4, pairs of scalar adjoints have couplings to pairs of gauge bosons through their kinetic terms.
However, couplings between a single adjoint and a pair of gauge bosons must come at higher dimension.
We will now consider the complete set of gauge and Lorentz invariant operators of dimension 5 where a scalar adjoint couples to pairs of gauge bosons.

For the scalar color octet the gauge and Lorentz invariant dimension 5 operators are

\begin{equation}
L= \frac{d^{abc}}{\Lambda_{2}} S_a G_b^{\mu \nu} G_{c, \mu \nu} + \frac{1}{\Lambda_{1}} S_{a} G^{a,\mu \nu} B_{ \mu \nu}
\end{equation}
where $\Lambda_i$ is the scale of new physics at which the operator is generated.  Here, $G^{\mu \nu}$ is the SU(3) field strength tensor and  $B_{ \mu \nu}$ is the U(1) field strength tensor.
The first operator couples the color adjoint to a pair of gluons, and is familiar from previous studies of color adjoint phenomenology.  
In this operator, the color indices of the gluons and adjoint scalar are contracted symmetrically with $d^{abc}$.
The second operator is also gauge and Lorentz invariant.  
In this operator, the Lorentz indices are contracted between field strength tensors, while the color index is contracted between the gluon and the color octet.
This is an effective coupling between the scalar octet, one gluon, and a photon/Z-boson.

Similarly we may consider a similar operator governing the SU(2) adjoint T

\begin{equation}
L= \frac{1}{\Lambda_{TWB}} T_i W_i^{\mu \nu} B_{ \mu \nu}
\end{equation}
where $W^{\mu \nu}$ is the SU(2) field strength tensor, and $B_{ \mu \nu}$ is the U(1) field strength tensor.
Lorentz indices are contracted between the field strength tensors, while the SU(2) indices are contracted between the adjoint T and the SU(2) field strength tensor.
Unlike SU(3), SU(2) does not have a factor $d_{abc}$, therefore we may not write a coupling between the field T and two  $W^{\mu \nu}$'s as above.
However, we may extend the list of possible gauge invariant operators if we allow the insertion of Higgs fields to soak up SU(2) indices.
These operators will be of dimension 7, but once Higgs vevs are inserted, they become operators of effective dimension 5.  These dimension 7 operators are

\begin{equation}
L= \frac{1}{\Lambda_{TBB}^3} [H^{\dagger} T H] B^{\mu \nu} B_{ \mu \nu}+ \frac{1}{\Lambda_{TWW^3}} [H^{\dagger} T H] W^{\mu \nu} W_{ \mu \nu} + \frac{1}{\Lambda_{TGG}^3} [H^{\dagger} T H] G^{\mu \nu} G_{ \mu \nu}+\frac{1}{\Lambda_{SGW}^3}  S^a G_a^{\mu \nu} [H^{\dagger}W^i_{ \mu \nu} H]
\end{equation}

In the first three terms we have contracted the SU(2) indices of the adjoint T with those of the Higgs fields to make an SU(2) singlet; this is indicated by brackets. In the last term, SU(2) indicies are contracted between the SU(2) field strength tensor  and the Higgs doublets.  
The SU(2) structure of these operators is quite reminiscent of loop induced operators in two Higgs doublet scenarios which couple the Higgs fields to SM field strength tensors as, for example, in references \cite{Mendez:1990epa}\cite{Arhrib:2006wd}.
The operators above are of dimension 7, however once the Higgs vev is inserted one get operators of effective dimension 5, which are suppressed by two powers of $\frac{v}{\Lambda}.$
In the appendix we make a more complete list of operators for more complicated models where the scalar color octet may have additional quantum numbers under the other SM gauge groups.
We expect the operators above may be generated at loop level by various UV completions of the theory. We give some illustrative examples in the next section.

\section{High Energy Models}
\begin{figure}[tbh-]
\centering
\includegraphics[width=.8\columnwidth]{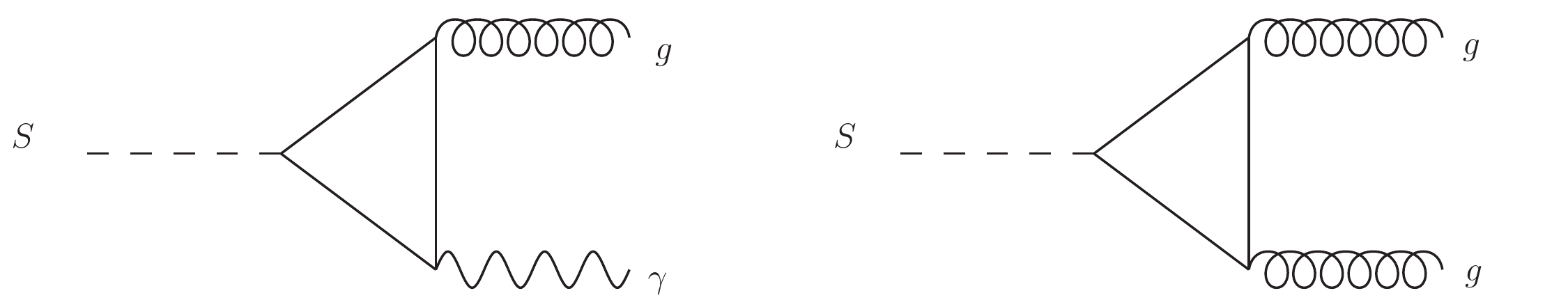}
\caption{Decay through mediators to di-boson final states.}
\label{Decay Loop}
\end{figure}

We expect our effective operators will be produced by integrating out heavy fields which have quantum numbers under several SM gauge groups.
One of the simplest completions may be to include one or more generations of heavy vector-like quarks or leptons which carry hypercharge in addition to SU(2) or SU(3) quantum numbers.
Sets of these vector-like "messenger" fields may couple to our scalar adjoint fields through Yukawa-like terms.
For example, a set of vector-like quarks which carry hypercharge have quantum numbers $(3,1)_Y$ and may couple to a scalar octet to generate the Feynmann diagrams in Fig. \ref{Decay Loop}.  
These diagrams correspond to the operators in Eqn 1.  The exact ratio of the operator coefficients $\Lambda_2$ and $\Lambda_1$ will depend on the hypercharge of the messengers.
A general estimate  for the ratio of effective couplings might be $g_{3}/g_{1}$, though this depends on the details of messenger sector. 
The effective coupling to U(1) gauge bosons is non-negligible and will provide additional production and decay channels for a scalar octet.  
This opens new windows for collider searches to illuminate new physics sectors.

One large class of models which contain SM adjoint scalars are supersymmetric theories in which gauginos are Dirac as opposed to Majorana \cite{Fox:2002bu}\cite{Hall:1990hq}.
In Dirac gaugino models, gauginos get mass by mixing with new chiral fields which are adjoints under the Standard Model gauge groups.  
In order to generate the Dirac Mass the relevant super-potential operator is
\begin{equation}
W= \int d^2 \theta \frac{W^{'}_{\alpha}W^{\alpha}A}{\Lambda} = \frac{D \lambda \psi_A}{\Lambda}
\end{equation}
where A denotes the chiral adjoint field, W is a standard model gauge field strength and $W'$ is the field strength of a hidden $U(1)$ gauge group which is broken at a high scale.  
This operator is expected to be generated by integrating out heavy messenger fields which are charged under the SM as well as the hidden-sector U(1) gauge group, thus the UV model is a form of gauge mediation \cite{Dine:1995ag}.
The hidden-sector U(1) field  gets a D-term vev, thus the operator becomes a Dirac mass for the SM gaugino, 'marrying' it to the fermionic piece of the chiral adjoint.
Defining $D/\Lambda \equiv m_D$ we find the operator in Equation 4 is equivalent to a Dirac mass term $m_D \lambda \psi_A$ for the gaugino.
The chiral multiplet A is complex, it thus contains both real and imaginary scalar fields which are adjoints under a SM gauge group.  
The masses of the real and imaginary parts of the scalar adjoint depend greatly on the details of the messenger sector of the model see for example, see for example \cite{Carpenter:2010as},\cite{Carpenter:2015mna},\cite{Csaki:2013fla}.
Many models lead to tachyonic masses for one or the other adjoint, and fixes may produce a great range of adjoint masses.  
If the scalar adjoints are lighter than the gauginos and sparticles in the theory, then we expect their decays to follow through loops to sets of SM gauge bosons or fermions.

We note that consistent with all symmetries in the theory we may also write the super-potential term

\begin{equation}
W=\int d^{2}\theta \frac{W_{\alpha}^{Y} W^{\alpha} A}{\Lambda}
\end{equation}
where A is the a chiral SM adjoint field, $W^Y$ is the hypercharge field strength and W is appropriate SU(2) or SU(3) field strength tensor.
This operator is nearly identical to the 'supersoft' operator which generates Dirac gaugino masses.    We may also write
\begin{equation}
W= \int d^{2}\theta\frac{d^{abc}W^a_3 W^b_3 S^c}{\Lambda^{'}}
\end{equation}
Where $W_3$ is the SU(3) field strength and S in the chiral SU(3) adjoint. Upon integration over $\theta$s, we see that these superpotential terms produce in the Lagrangian our operators from Eqns 1 and 2. 
These operators are presumably generated much the same way as the supersoft term is, by integrating heavy fields out of the theory as we discuss below. 
We note that the Lagrangian terms in Eqns 1 and involve the real part of the adjoint.  We might also expect similar operators generated by messenger loops to couple imaginary part of A and the SM field strength tensor and dual.

Many aspects of the phenomenology of scalar adjoints resultant from R symmetric SUSY models have been studied thoroughly \cite{Plehn:2008ae}\cite{Choi:2008ub}.  
It is known, for example, that an octet-gluon-gluon coupling may be generated by integrating out squarks from the R symmetric theory. 
Similar diagrams involving squarks and sleptons will generate the the operators in Eqn 5. 
Squarks and sleptons couple to the real part of the chiral adjoints through Kahler potential D-terms. Due to a cancelation of diagrams, it is known that operators following from sparticle loops are suppressed when left and right sparticle masses are degenerate.

Another source of  operators in Eqn 5 and 6 are loops of the heavy messenger fields.
These messengers carry SM quantum numbers, and in many completions are  often various sets of fundamentals and anti-fundamentals under SU(5), see for example \cite{Carpenter:2015mna},\cite{Csaki:2013fla}. The relative size of the operators in Eqns 5 and 6 will depend on the details of the messenger sector, and on the sparticle spectrum.

Effective dimension 7 operators like those of Eqn 3 may be produced through loops of squarks and sleptons, as these sparticles are charged under multiple SM gauge groups and couple to the Higgs.  
Thus in 'supersoft' models we find numerous effective operators which couple scalar adjoints to pairs of gauge bosons.
As a phenomenological note, the real part of the supersoft adjoints may also decay to quarks/leptons through loops.
These decays, however, do not always dominate adjoint branching fractions - there are various regions of parameter space where these decays are subdominant to decays into gauge bosons.  
This region of parameter space will be of special interest to the phenomenological discussions of the next section.

\section{Production and Decay of Scalar Adjoints}

We will now consider the collider phenomenology of our effective operators.  We will focus on the phenomenology of colored states as they should be produced with large cross section at the LHC.
As such, we will study the production and decay of a single scalar color octet.
We will assume that, besides mass and kinetic terms, the octet has only dimension 5 couplings to SM gauge bosons resulting from the operators in Eqn 1. Our simple model has three parameters, the mass of the octet $m_{S}$, and the two effective operator coefficients $\Lambda_{1}$ and $\Lambda_{2}$.  In a UV complete model we would know the exact ratio between the operator coefficients, not knowing the details of the high energy model, however, we will consider these two parameters to be independent. The two operator coefficients control the couplings of the scalar color octet to three pairs of SM gauge bosons $gg$ and $g\gamma$ and  $gZ$.

We consider here only the single octet production process $p p > S $, which proceeds through gluon fusion.
The total production cross section is proportional to the gluon-gluon decay width, $\Gamma_{gg}$, of the octet, and is given by,

\begin{equation}
\sigma(pp\rightarrow S)=\Gamma_{gg} \varepsilon \frac{16\pi^{2}}{sm_{S}}\int_{\frac{m_{S}^2}{s}}^{1}\frac{dx}{x}f_{g}(x)f_{g}(\frac{m_{S}^{2}}{sx})
\end{equation}
where $m_{S}$ is the mass of the scalar, $f_{g}$ are the parton distribution functions (PDFs) for gluons, $x$ is the momentum fraction of the initial gluons, and $\varepsilon=1/32$ takes into account the interchange of summed and averaged over states in the decay rate and production cross section. The octet decay width into gluons $\Gamma_{gg}$ is given by
\begin{equation}
\Gamma_{gg}=\frac{40}{3}\frac{m_{S}^{3}}{32 \pi \Lambda_{2}^{2}}.
\end{equation}
We note that the decay width scales like $m_{S}^{3}$ leading to fairly wide resonances for larger masses of the octet.

The production cross section of the octet decreases with increasing parameter $m_S$, it also decreases with increasing scale of the effective cut-off $\Lambda_2$.
Higher order corrections to the production of color-octet scalars can have very large effects \cite{Idilbi:2009cc}. A K-factor is used to scale up the tree level production cross section of the scalar octet. A typical K-factor for the process $gg\rightarrow S$ at $\sqrt{s}=14$ is between 2.4 and 3.6 for a TeV scale octet.  In the analyses in this work we choose a K-factor of 2 for octet production at $\sqrt{s}=8$ TeV and a K-factor of 2.5 for a sensitivity projection at $\sqrt{s}=14$ TeV.

\begin{figure}[tbh-]
\centering
\includegraphics[width=.65\columnwidth]{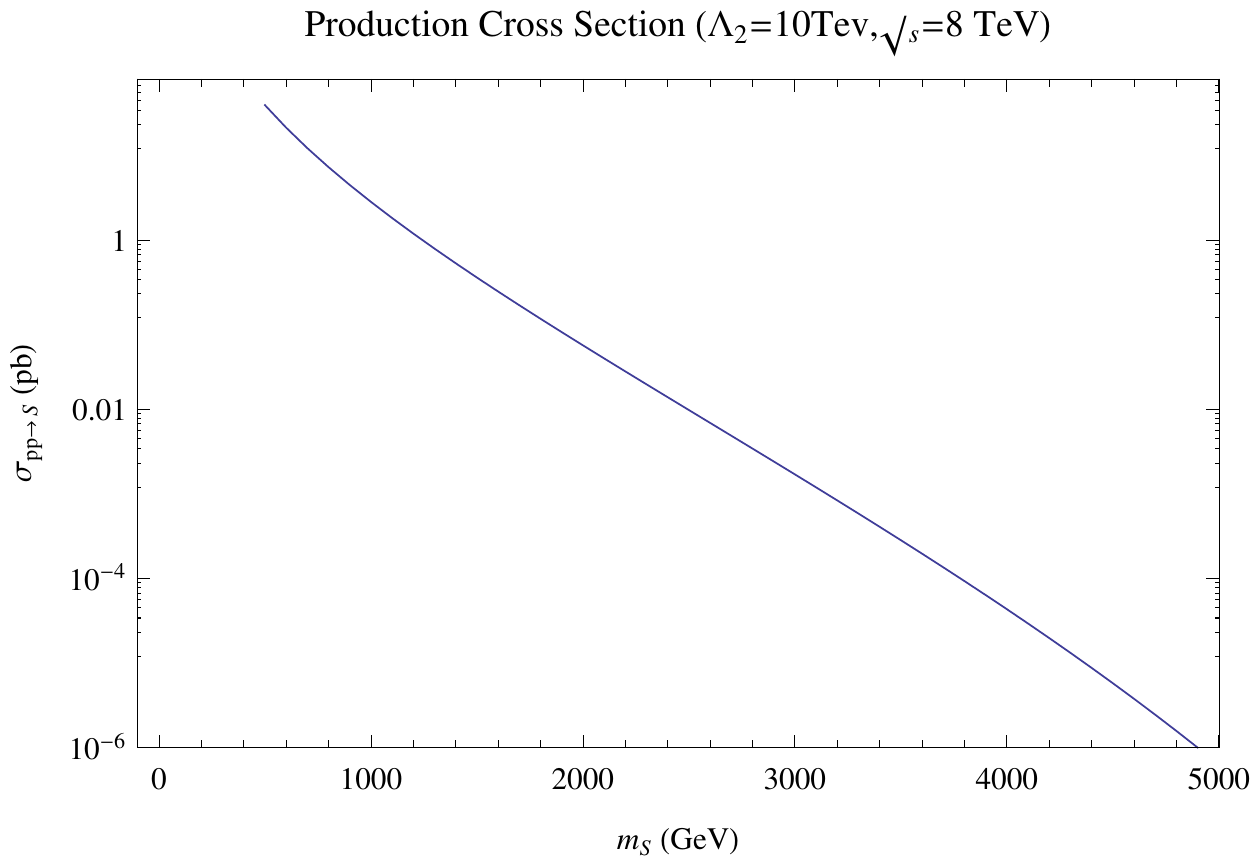}
\caption{Production cross section of a single scalar octet vs. $m_{S}$ in pp collisions at $\sqrt{s}=8$ TeV with $\Lambda_{2}=10$ TeV.  These cross sections were generated with MadGraph 5.}
\label{Single Production xsection}
\end{figure}

In Figure \ref{Single Production xsection} we plot the octet production cross section vs. $m_{S}$ for the process $gg\rightarrow S$ in 8 TeV proton-proton collisions.  In the figure we have chosen the effective operator coefficient $\Lambda_2$ to be 10 TeV, but this total production cross section may easily be re-scaled for various values of the effective cut-off.

We will now consider the decays of the scalar octet. In our model there are three possible di-boson decay channels for the octet ,$S\rightarrow gg, ~g\gamma, ~gZ$ .
That is, decays into two gluons, a gluon photon pair, and a gluon Z pair.  Using the effective Lagrangian in Eqn 1, we may calculate the branching fractions.  In the limit that the octet is much heavier than the Z boson, that is $m_Z<<m_S$, the branching fractions are given by,

\begin{equation}
\begin{split}
&BR_{g\gamma}=\frac{\Gamma_{g\gamma}}{\Gamma_{g\gamma}+\Gamma_{gg}+\Gamma_{gZ}}=\frac{c_{W}^{2}}{1+\frac{10}{3}\Lambda_{1}^{2}/\Lambda_{2}^{2}} \\
&BR_{gZ}=\frac{\Gamma_{gZ}}{\Gamma_{g\gamma}+\Gamma_{gg}+\Gamma_{gZ}}=\frac{s_{W}^{2}}{1+\frac{10}{3}\Lambda_{1}^{2}/\Lambda_{2}^{2}} \\
&BR_{gg}=\frac{\Gamma_{gg}}{\Gamma_{g\gamma}+\Gamma_{gg}+\Gamma_{gZ}}
=\frac{1}{1+\frac{3}{10}\Lambda_{2}^{2}/\Lambda_{1}^{2}}
\end{split}
\end{equation}
where $s_{W}$ and $c_{W}$ are the sine and cosine of the weak angle and $\Lambda_i$ are the effective operator coefficients.  
The exact form of the branching fractions are given in the Appendix.  
By varying the ratio of effective couplings $\Lambda_1/\Lambda_2$, we may change the branching fraction rates into the various diboson channels.  
For generic completions of the effective theory, for example including sets heavy vector-like quarks with electric charge in the range 1/3, we expect the ratio of the branching fractions  $\Gamma_{g\gamma}/\Gamma_{gg}$ to be in the percent range.  This branching fraction is not insignificant, and together with the large octet production cross section  it will mean that there is a substantial signal in the $g\gamma$ resonance channel.  As we will see below, this is an interesting and a relatively clean channel, it will therefore be important for scalar octet searches.

\section{Current Limits}

We will now discuss the bounds on parameter space which result from various 8 TeV analyses in Run I of the LHC.  
In our model, production and decay of the octet produces many interesting final state topologies, therefore there are multiple search channels that can be relevant in the constraint or discovery of the scenario. 
Both ATLAS and CMS have performed searches for heavy resonances decaying into various final states, and one standard search for new colored resonances is in the dijet resonance channel, $pp \rightarrow X \rightarrow j j$ \cite{Aad:2014aqa}\cite{Khachatryan:2015sja}. 
Here a heavy state, $X$, is produced and then decays hadronically.
These searches constrain the parameter space of our color adjoint operators as the scalar octet generically has significant branching fraction into the gluon-gluon final state.
In addition ATLAS and CMS have performed searches for resonances which decay to a photon and jet final state \cite{Aad:2013cva} \cite{Khachatryan:2014aka}.  
This search is relevant to our model since the scalar has a decay mode $S\rightarrow g\gamma$.
The monojet search $p p \rightarrow j+\met$ is an important inclusive search channel for constraining new physics \cite{Aad:2015zva}\cite{Khachatryan:2014rra}.  
This analysis will be relevant to our decay channel $S\rightarrow gZ$, where the Z decays to missing energy $(Z\rightarrow \nu\nu)$.
Finally, CMS has a search for a heavy resonance decaying to a jet and a hadronically tagged vector boson $pp\rightarrow X \rightarrow j+V$ \cite{Khachatryan:2014hpa}. 
This analysis will be relevant to our decay channel $S\rightarrow gZ$, where the Z decays hadronically $(Z\rightarrow j j )$.  
Below, we will explore the bounds on our parameter space from each of these studies.

\subsection{The Jet + $\gamma$ Channel}
We first discuss constraints from searches for $\gamma + j$ resonances on the  process  $gg \rightarrow S \rightarrow \gamma g$.
The benchmark models used in these analyses include production of  non-thermal quantum black holes (QBH) and excited quarks ($q^{*}$), however the search constrains our color octet model as well.

The ATLAS search for photon plus jets used the following search criteria:

\begin{itemize}
 \item At least one isolated photon with $p_{T}>125$ GeV
 \item At least one hard jet with $ p_{T}> 125$ GeV.
 \item Photon required to have angular separation $\Delta R(\gamma , j) > 1.0$ between leading photon and all other jets with $p_{T}>30$ GeV.
 \item Photon and jet required to be in the central region of the detector with $|\eta_{\gamma}|< 1.37$ and $|\eta_{j}|< 2.8$.
 \item Pseudo-rapidity separation between jet and photon of $|\Delta \eta (\gamma , j)| < 1.6 $.
 \item Highest $p_{T}$ $\gamma$ and jet candidates used to compute $m_{\gamma j}$, which is binned.
\end{itemize}

\begin{figure}[tbh-]
\centering
\includegraphics[width=.65\columnwidth]{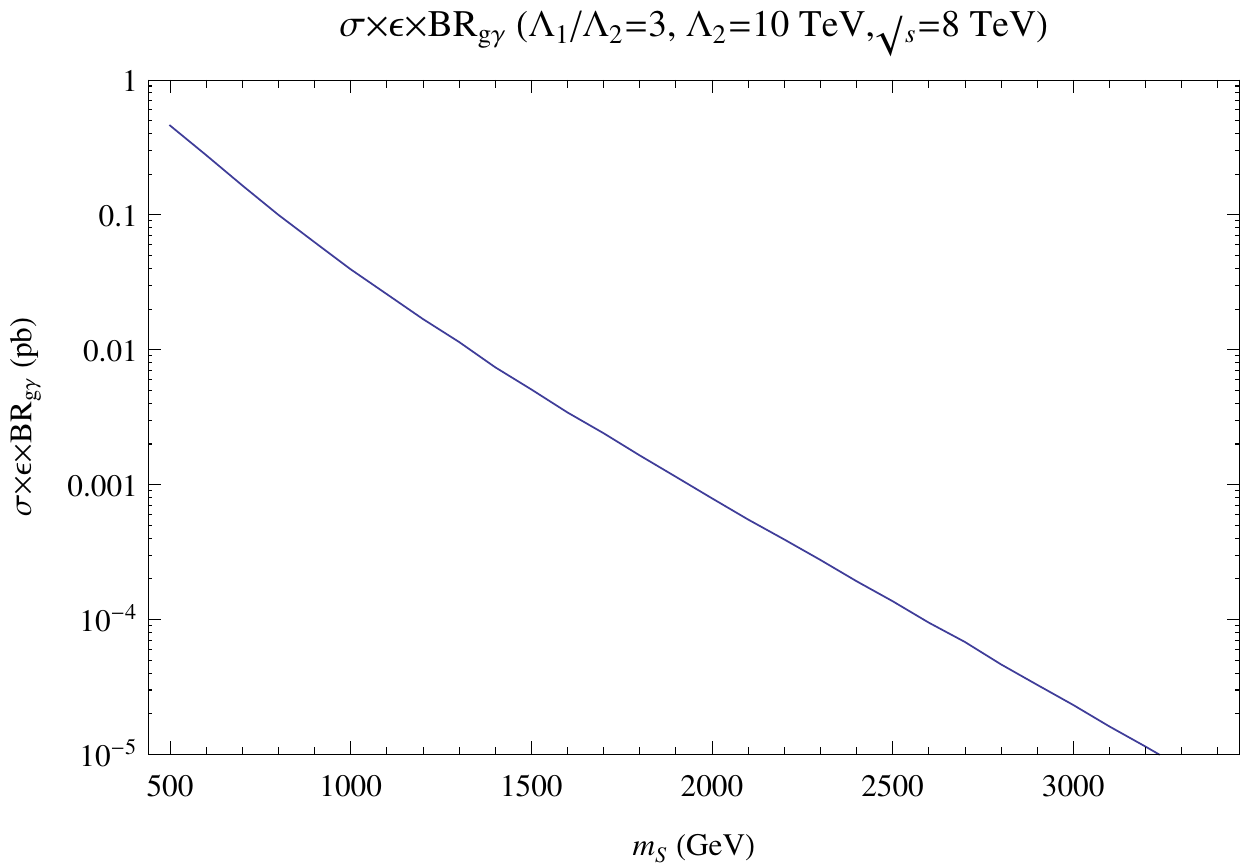}
\caption{The rate $\sigma\times\epsilon\times BR_{g\gamma}$ of the scalar octet resonance at 8 TeV where $\Lambda_{2}=10$ TeV and$\Lambda_{1}=30$ TeV. This rate includes the efficiency with which events passes the ATLAS selection criteria.}
\label{effic_xsec}
\end{figure}

The SM background for this search is dominated by $\gamma + j$ production and QCD dijet production.
The SM production of $\gamma + j$ lacks a resonant process and is generated via Compton scattering of quarks and gluons, and through quark-antiquark annihilations.
Electroweak processes account for about 1\% of background.
The analysis performed by ATLAS observed no excess of events over SM backgrounds, and reported limits on production and decay of their benchmark models to $\gamma j$ final states.

We recast the ATLAS search at $\sqrt{s}=8$ TeV with $20.3$ fb$^{-1}$ of data in order to rule on the parameter space of our color octet model.
We implemented the effective Lagrangian using FeynRules \cite{Alloul:2013bka}, then using MADGRAPH 5 \cite{Alwall:2011uj} we generated events for the process $gg\rightarrow S\rightarrow \gamma g$ in pp collisions at 8 TeV. The events were showered with Pythia \cite{Sjostrand:2006za} and run through the PGS detector simulator \cite{PGS}.
We then implemented the ATLAS cut-based analyses on our data to determine the efficiency of the ATLAS search for our model.
A plot of signal efficiency vs $m_S$ is given in Figure 4a.

In general, we find signal  efficiencies of order 50 percent for scalar masses between 1 and and 4 TeV, with the efficiency dropping rapidly at low invariant mass.

For each point in parameter space we specify the values of  $ m_S$, $\Lambda_{1}$, $\Lambda_{2}$. 
We then calculate the total number of events which pass the selection criteria, $N_{evt}= \sigma \times L\times\epsilon \times BR_{g\gamma}$, in each invariant mass bin.  In Figure 3 we plot the total rate $\sigma \times\epsilon \times BR_{g\gamma}$ vs. scalar mass $m_S$ for the benchmark values $\Lambda_2 =$ 10TeV and $\Lambda_1=$ 30 TeV.  
At this benchmark point, for octet masses below about 2 TeV, the total rate is in the few fb range, and therefore potentially detectable.
We show the 95\% confidence level (CL) exclusions in the $\Lambda_{1}$-$\Lambda_{2}$ plane for various values of the scalar mass in Fig. \ref{fig:eftbounds}.

We see that exclusions become weaker as the scalar mass increases, as is expected due to a rapidly decreasing production cross section.  
For a given mass, the production rate is set by the parameter $\Lambda_2$, with the rate decreasing as $\Lambda_2$ increases.  
The branching fraction into the $\g\gamma$ final state is set by the ratio of scales $\Lambda_1/ \Lambda_2$.  
Thus the total signal rate decreases as this ratio of scales increases.  
We may exclude effective cut-offs in the 5-10's of TeV range for $m_{S}$ of the order a few TeV.  
We note that the effective operator paradigm is not valid to arbitrarily low cut-off scales, but will be valid only when the $\sqrt{s}<2 M_{mess}$ where $M_{mess}$ are the mass of the UV messengers integrated out of the theory.

\subsection{Dijet Channel}

Dijet searches tightly constrain the existence of new colored resonances.
The ATLAS search for dijet resonances was done at $\sqrt{s}=8$ TeV with $20.3$ fb$^{-1}$ of data.
The benchmark models that are considered by ATLAS include excited quarks, QBHs, heavy $W'$ gauge fields decaying to quarks, and scalar octet particles decaying to a pair of gluons.

The ATLAS dijet search criteria is as follows:

\begin{itemize}
 \item At least 2 anti-kt jets with rapidity $|y|<2.8$.
 \item Two hardest jets (leading and subleading) are required to have  $p_T>50$ GeV.
 \item An invariant mass cut of $m_{jj} > 250$ GeV is placed on the two hardest jets
 \item Separation of rapidities $\frac{1}{2}|y_{\text{lead}}-y_{\text{sublead}}|<0.6$.
\end{itemize}

SM backgrounds in this channel are dominated by QCD processes with a sub-percent mixture of additional SM processes. The background has a smoothly falling spectrum as the dijet invariant mass of the two highest $p_{T}$ jets (leading and sub-leading), $m_{jj}$ increases \cite{Harris:2011bh}.

ATLAS analyzed these results to constrain a scalar octet benchmark model. Efficiency cuts for this model range from 61\% to 63\%. The ATLAS analysis showed no excess for dijet invariant masses up to 4.5 TeV and placed limits on production rates of their octet.  The 95\% CL upper-limit on $\sigma\times A$ with 100\% branching fraction to $gg$ final state was reported by ATLAS.  We  recast these results to set bounds on the parameter space of model in which the scalar octet has multiple decay channels.  By rescaling the ATLAS total rate for the process $gg\rightarrow S \rightarrow gg$ we place bounds in the $\Lambda_1 -\Lambda_2$ plane for various values of $m_{S}$.  We show exclusions in Fig. 5b.

For a given octet mass $m_S$, the total production rate depends on the effective coefficient $\Lambda_2$, with the rate decreasing as $\Lambda_2$ increases.  The total branching fraction of the octet into gluons is then given by the ratio of scales $\Lambda_1/\Lambda_2$, with the di-gluon rate increasing as this ratio is increased. We thus expect this search to put the tightest limits on the scalar octet model when $\Lambda_{1}$ is large compared to $\Lambda_{2}$, that is, when the effective couplings $Sg\gamma$ and $SgZ$ are suppressed.

\subsection{The Monojet Channel }
The mono-jet search channel is a standard inclusive analysis that applies to many BSM scenarios.
 In the model we are considering, the process $gg\rightarrow S \rightarrow Z g$ will lead to a host of interesting signal topologies, among which is a monojet signature.  This final state is achieved when the $Z$ decays invisibly.
We will now recast the ATLAS $\sqrt{s}=8$ TeV, $20.3$ fb$^{-1}$ monojet analysis to rule on our scenario.

Monojet searches look for an event topology with a hard jet in the central detector region and several hundred GeV of missing energy.
The ATLAS monojet search selection criteria is as follows:

\begin{itemize}
 \item Events required to have $\met>150$ GeV and at least one jet of $p_T>30$ GeV with $|\eta| < 4.5$.
 \item Leading jet with $p_T>120$ GeV and $|\eta|<2.0$.
 \item For a monojet like topology, the ratio $ p_T/\met>0.5$.
 \item Separation between leading jet and direction of missing transverse momentum $\Delta\phi (j, \met)>1.0$.
 \item Veto of events with electrons or muons with $p_T>7$ GeV or isolated tracks with $p_T>10$ GeV and $|\eta|<2.5$.
\end{itemize}

Events passing the cuts were divided into 9 signal regions (SR) defined by the minimum amount of $\met$ in the SR.
Limits where 95\% CL limits were placed on $\sigma\times A\times\epsilon$ for the different SR as no excess of events was observed.
The signal regions for the ATLAS analysis are shown in the table below.

\vspace{2mm}
\begin{center}
\begin{tabular}{| c | c | c | c | c | c | c | c | c | c |}
  \hline
  Signal Regions& SR1 & SR2 & SR3 & SR4 & SR5 & SR6 & SR7 & SR8 & SR9 \\
  \hline
  Minimum $\met$ (GeV) & 150 & 200 & 250 & 300 & 350 & 400 & 500 & 600 & 700 \\
  \hline
  95\% CL exclusion for $\sigma\times A\times\epsilon$ (fb) & 726 & 194 & 90 & 45 & 21 & 12 & 7.2 & 3.8 & 3.4\\
  \hline
\end{tabular}
\end{center}
\vspace{2mm}

The main backgrounds for this search are dominated by $Z(\rightarrow \nu\nu)$+jets and $W$+jets production. Smaller contributions come from $Z/\gamma^{*}(\rightarrow l^{+}l^{-})$+jets, multijet, $t\bar{t}$, and diboson processes.
The search found no excess of monojet events in the 8 TeV dataset.

We utilize this search results to constrain parameter space.
We generate events for the process $pp\rightarrow S \rightarrow Z g \rightarrow \nu\nu g$ in MADGRAPH 5. The events are showered with Pythia and then run through the PGS detector simulator.
We then implemented the mono-jet cut based analysis to calculate the efficiency for events in the various signal regions.  
The calculated efficiency for scalar octets model in various signal regions is found in Fig 4b. 
The overall efficiency of the mono-jet search for this process is fairly high. We note that our events differ from many models searched for in the monojet channel.  For example a cast of DM models in which a gluon is emitted as initial state radiation favor softer gluons with $p_T$ peaked at low values. In our events however, the gluon in emitted as a hard decay product in the final state, thus will have a much harder $p_T$ which is more amenable to passing a hard $p_T$ cut.

We then calculate the total expected event rate and compare to the reported 95\% CL exclusion limits.  The exclusions limit results in the $\Lambda_{1}$-$\Lambda_{2}$ plane can be seen in Fig. 5c for various values of the octet mass $m_S$.
The mono-jet search does significantly worse at constraining the scalar octet model than the dijet or $\gamma j$ searches. 
This is in part because the multiplicative effects of the octet branching ratios into the $gZ$ channel and the Z branching fraction into neutrinos. 

As before, the overall production rate of octets decreases with increasing scale $\Lambda_2$. 
The branching fraction into the $gZ$ final state decreases as the ratio $\Lambda_1/\Lambda_2$ increases. 
Thus the shape of the exclusion due to this search is similar to that from the $g\gamma$ search.   
We note that the ratio of $g \gamma$ to $g Z$ events is strictly related by gauge invariance given the form of our operator with coefficient $\Lambda_{1}$, therefore we expect that any signal for scalar octets observed in the $g\gamma$ would eventually produce a signal in the $gZ$ channel as well.

\subsection{Heavy Boson plus Jet Channel}

The CMS collaboration has searched for heavy resonances which decay to a jet and a massive vector boson in the process $pp\rightarrow X \rightarrow j+V$. 
The search also applies to heavy resonances which decay to two heavy vector bosons.
In these searches the vector bosons decay hadronically, $V\rightarrow j j $, and is reconstructed with a hadronic tag.
Benchmark models considered for the CMS analysis are excited quarks ($q^{*}\rightarrow q W$, $q^{*}\rightarrow q Z$), Randall-Sundrum gravitons ($G_{RS}\rightarrow WW$), and heavy $W'$ bosons ($W'\rightarrow WZ$).  The search also applies to our scenario where a heavy scalar octet decays to a gluon and hadronically decaying Z boson, $gg\rightarrow S\rightarrow Z g, Z\rightarrow j j$.

To identify a jet as being Z/W-tagged, the CMS analysis used jet pruning techniques, and require the pruned "fat-jet" with to have mass between   $70~$GeV$<m_{j}<100~$GeV.
The search criteria for the CMS analysis is as follows:

\begin{itemize}
 \item One vector-boson tagged "fat"-jet
 \item At least two jets with $p_{T}^{jet} > 30~$GeV and $|\eta|< 2.5$.
 \item At least one of the two highest $p_{T}$ jets is required to be Z-tagged.
 \item The pseudorapidity separation between the leading jets $|\Delta \eta (j,j)|<1.3$.
 \item A cut on the invariant mass of the two highest $p_{T}$ jets of $m_{jj}>890~$ GeV
\end{itemize}

\begin{figure}[tbh-]
\centering
\includegraphics[scale=0.64]{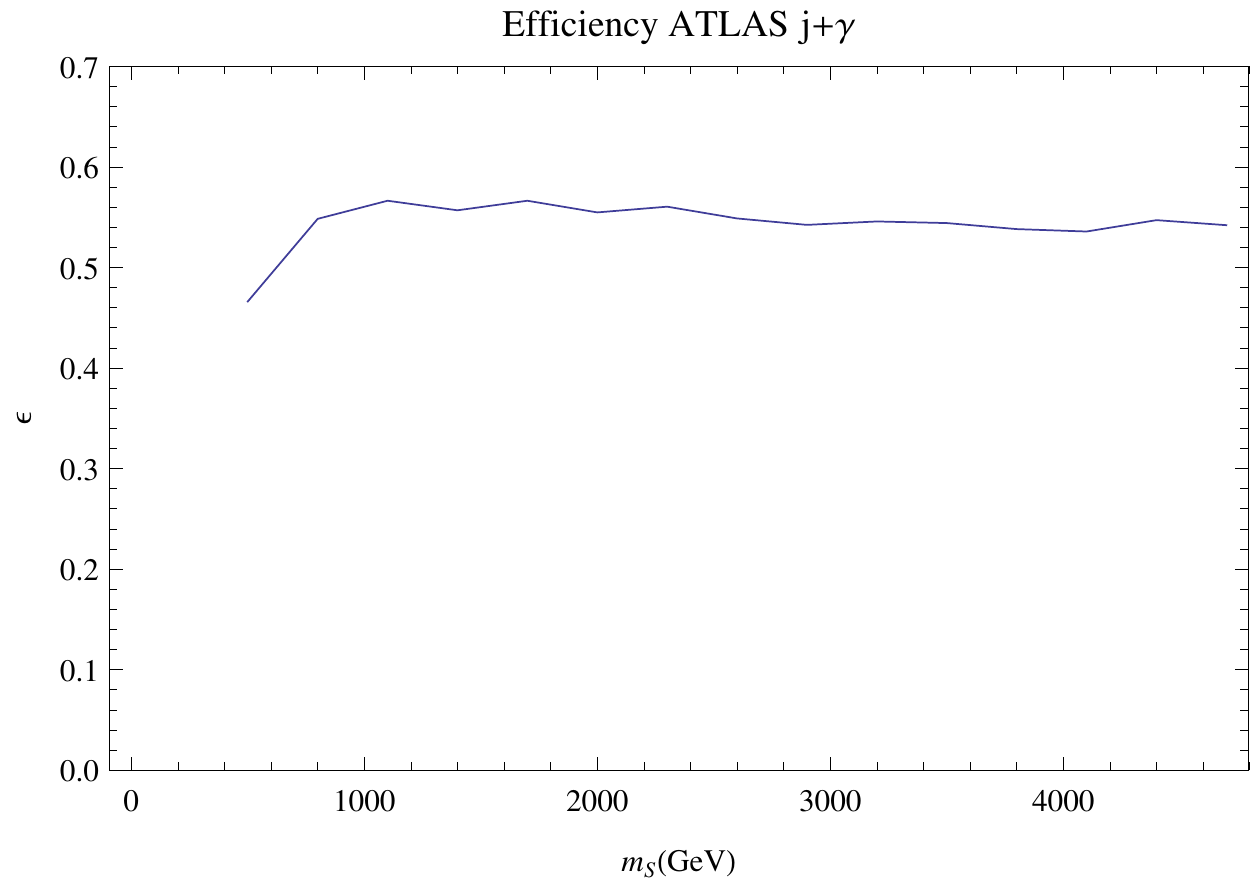}
\includegraphics[scale=0.90]{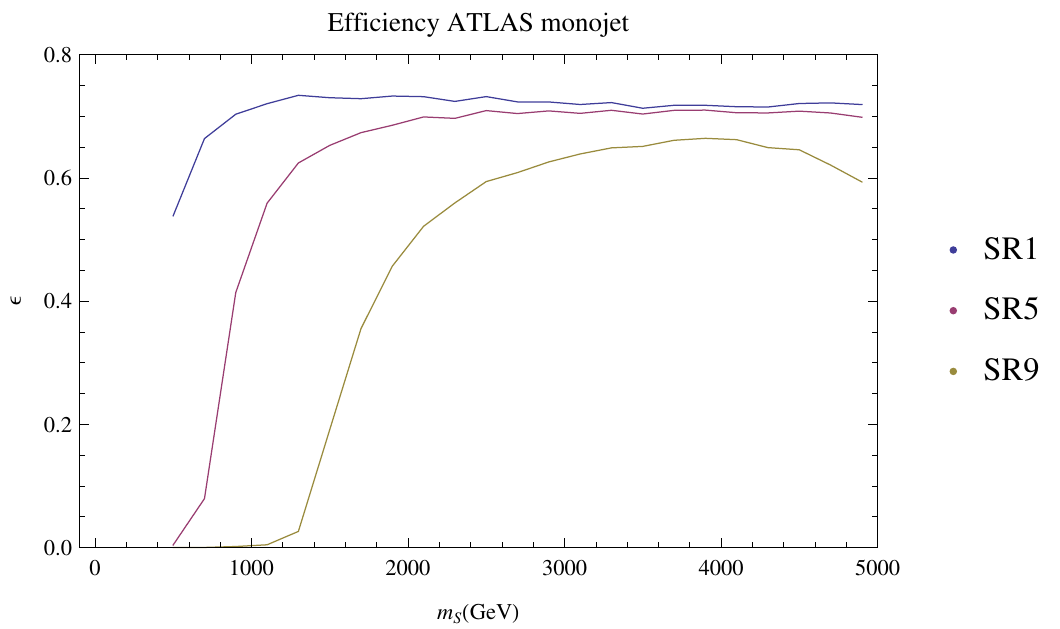}
\includegraphics[scale=0.64]{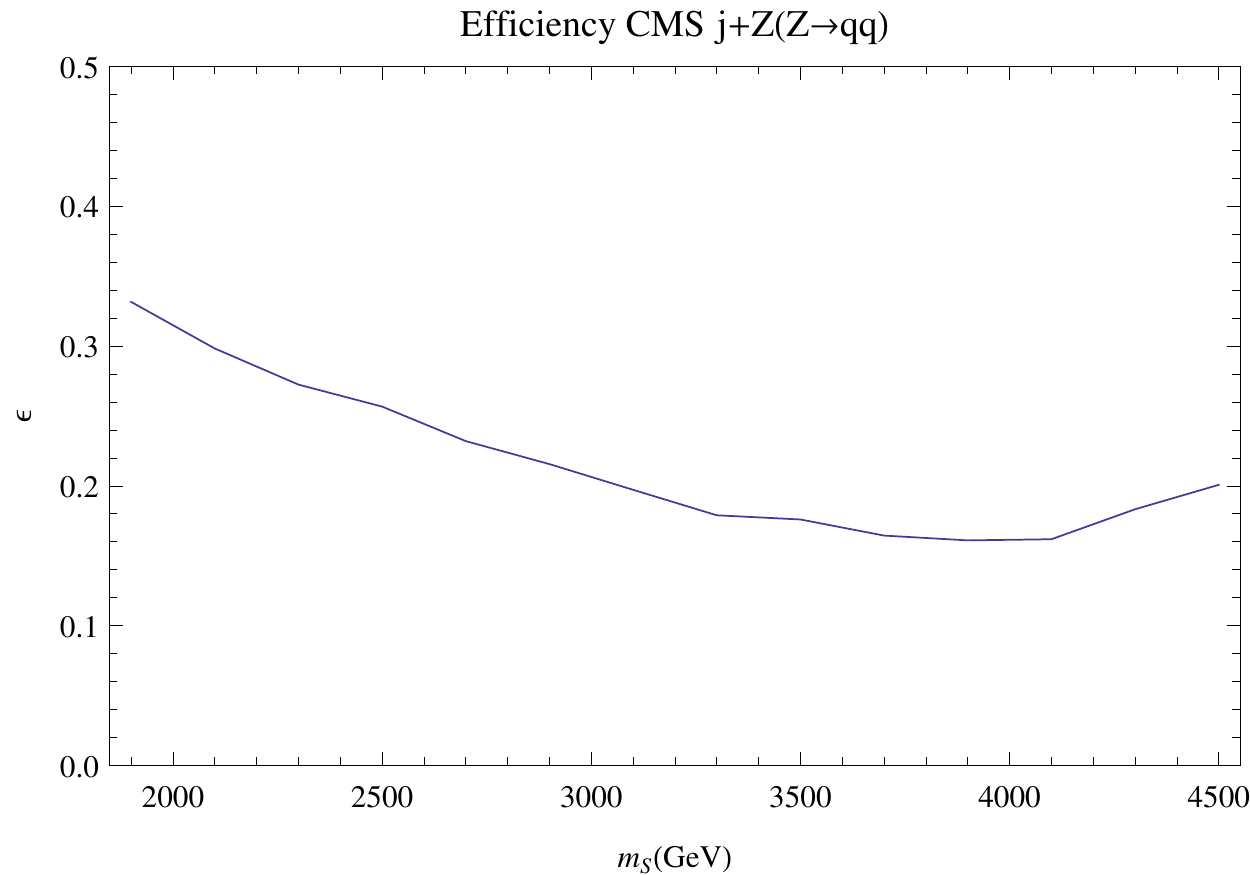}
\caption{Efficiency of our simulated events passing the selection criteria in the search channels considered for $\sqrt{s}=8$ TeV analyses at the LHC.
The upper plots are the signal efficiencies for the ATLAS searches for photon plus jet (left) and monojets (right) corresponding to the decay modes $S\rightarrow g\gamma$ and $S\rightarrow gZ(Z\rightarrow \nu\nu)$ respectively.
The lower plot is the kinematic efficiency of the CMS search for hadronically decaying heavy bosons for the decay mode $S\rightarrow gZ(Z\rightarrow qq)$.}
\label{fig:efficiencies}
\end{figure}

The dominant source of background for this search stems from multijet production with $t\bar{t}$, $W$+jets, and $Z$+jets contributing less than 2\%.
Data was binned by the invariant mass of the jet and tagged "fat"-jet $m_{jj}$. 
No significant excess in data was reported.  We may thus use this search to constrain our parameter space.

Using MADGRAPH 5 we simulated parton level events events for the process  $gg\rightarrow S\rightarrow Z g, Z\rightarrow j j$.  We estimated the search efficiency for these events using the CMS "fat"jet-tagging efficiency and cuts for kinematic efficiency.  The tagging efficiency reported by CMS for a single Z-tagged fat-jet ranges from roughly 45\% to 30\%, with the Z-tag doing better at lower invariant mass.  In our analysis we took the Z-tag jet efficiency to be 40\%.  In Figure 4c we plot the search efficiency vs octet mass $m_S$ in the Z+j resonant final state.

Using the production cross section, branching fraction and efficiency for our events, we may then calculate the total number of expected events in each invariant mass bin for each point in our parameter space. We then compare to 95\% CL upper bounds in each mass bin from the CMS search to constrain the parameters.  Exclusions are shown in the $\Lambda_1-\Lambda_2$ plane for various scalar masses $m_S$ in Fig. 5d.  The shape of the exclusion line follows from previous discussions of the $g\gamma$ and $g Z$ final states.

\subsection{Combined Results}

\begin{figure}[tbh-]
\centering
\includegraphics[scale=0.74]{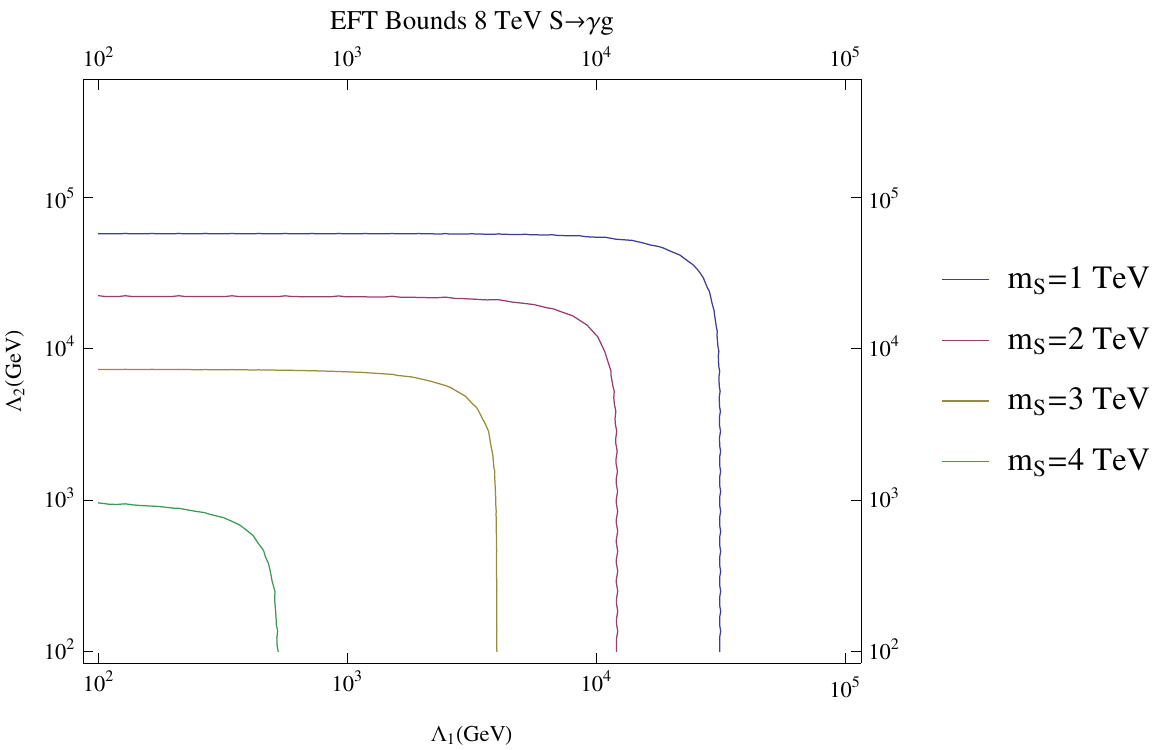}
\includegraphics[scale=0.74]{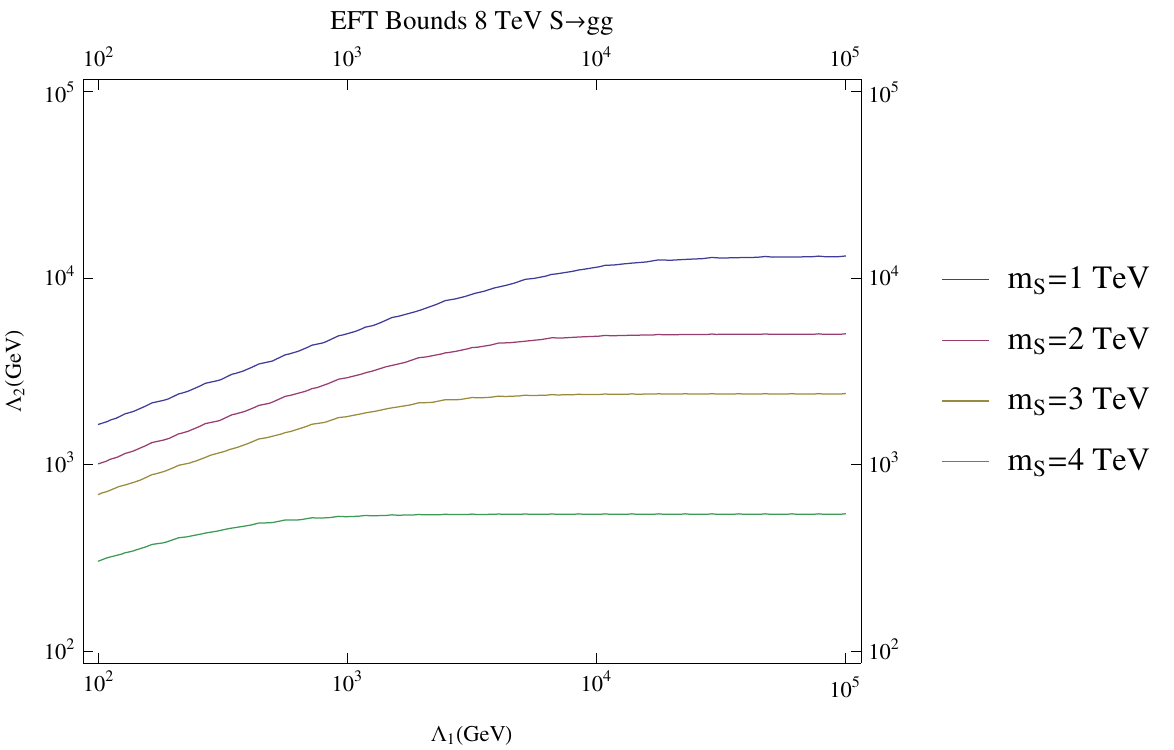}
\includegraphics[scale=0.74]{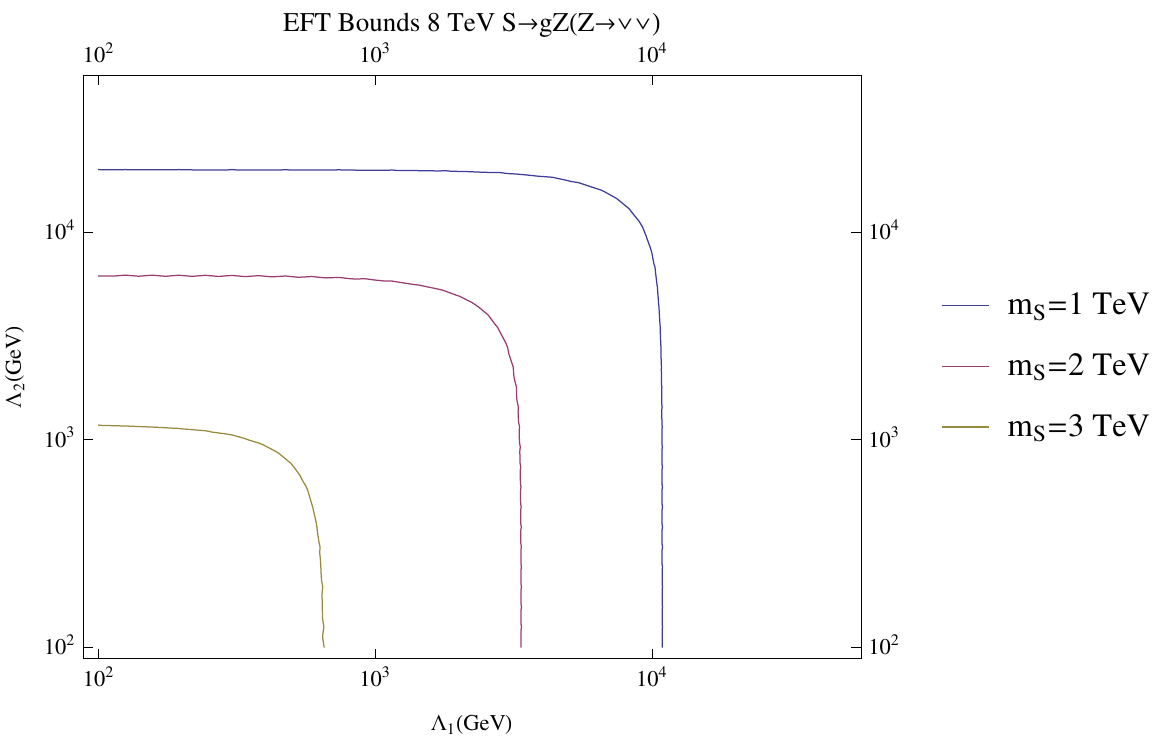}
\includegraphics[scale=0.74]{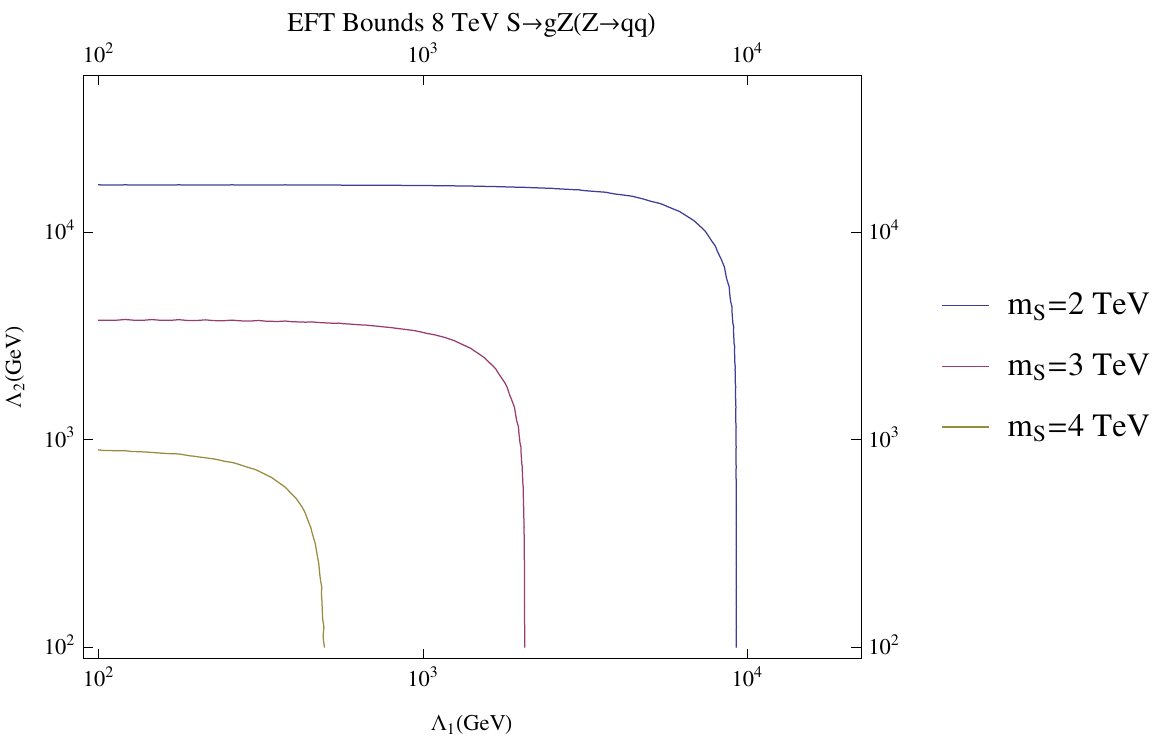}
\caption{Bounds on the $\Lambda_1-\Lambda_2$ for 4 values of the octer mass $m_S$.  Exclusions follow from 4 LHC 8 TeV analyses.  The top left follow from searches in the $pp\rightarrow X\rightarrow g\gamma $ final state.   The top right follow from searches in the $pp\rightarrow X\rightarrow j j $ final state.  The bottom left follow from searches in the $pp\rightarrow  j + \met $ final state.  The bottom right follow from searches in the $pp\rightarrow j+V, V\rightarrow jj$ final state. }
\label{fig:eftbounds}
\end{figure}

\begin{figure}[tbh-]
\centering
\includegraphics[width=.7\columnwidth]{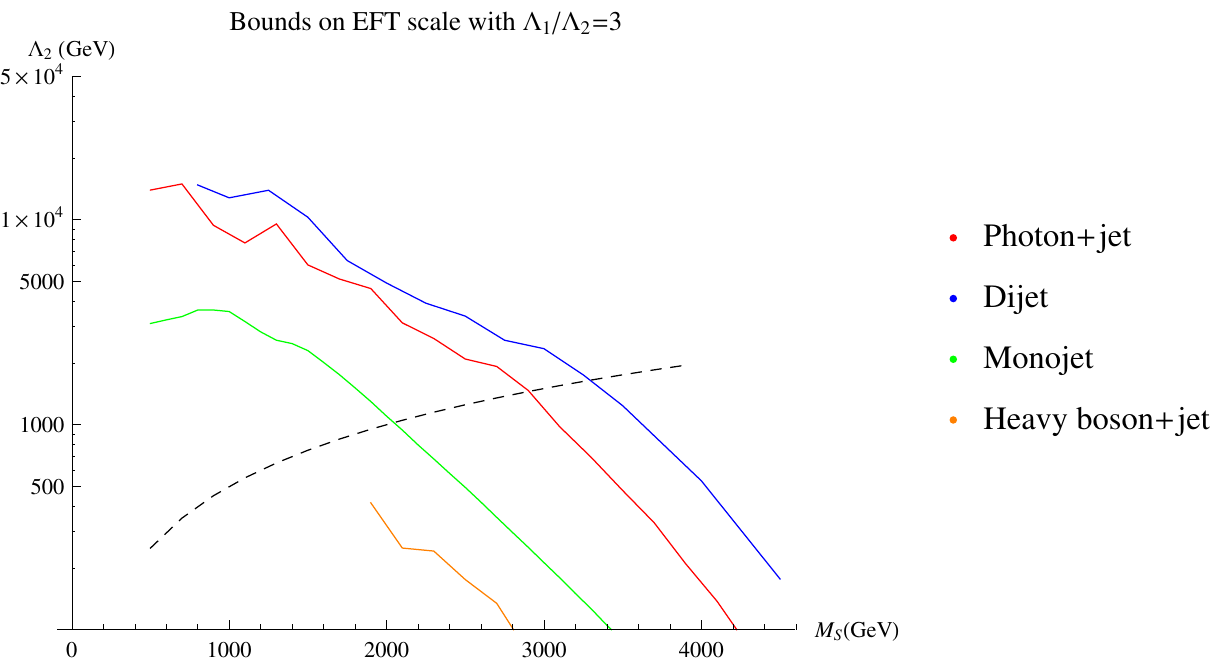}
\caption{Combined 95\% confidence level bound on EFT scale from all the channels considered for the choice $\Lambda_{2}=.1\Lambda_{1}$.  The black dashed line corresponds to where the validity of the EFT framework breaks down $m_{S}>2\Lambda_{i}$.}
\label{cutvmass}
\end{figure}

Combined results for all of the  final state channels under consideration are displayed in Fig. \ref{cutvmass}.  In the figure we have fixed the scale $\Lambda_2$ to be 10 TeV, and the ratio  $\Lambda_{1}/\Lambda_{2}$ to be 3. This benchmark point is representative of a class of a UV completions with vector-like messengers in the multi-TeV range\footnote{We implemented a model in Feynrules containing a set of heavy vector-like quarks with qunatum numbers $(3,1)_{2/3}$ and a scalar octet.  The model was exported to Madgraph NLO to calculate branching fractions of the octet.  Resultant branching fractions confirm the benchmark ratio of effective coefficients}.  Effective cutoffs may be excluded between a few and 10s of TeV over the octet mass range.
We see that for high $m_S$ regions the dijet search is the most exclusive, while the $\gamma +j$ search is the most exclusive for the low $m_S$ regions.  While the branching fraction of the octet into $\gamma +j$ is in the percent range for this benchmark point, the background for this channel is quite low.  Thus we find that in some of the parameter space of the theory, searches in the $\gamma +j$ channel may be competitive with dijet searches.  Due to a very large background, the dijet channel signal efficiency for very light octets is quite low.  We might expect that in searches for scalar octets with masses in the 100s of GeV range, $\gamma +j$ will generally be the most powerful.

We will now discuss the limits of the effective operator paradigm. The effective operators which we have been studying are generated by integrating out some heavy messenger particles which couple to the scalar octet and to gauge bosons. We know that the effective operator treatment is invalid for scales of the effective cut-off that are below the center of mass energy of the process, that is where the messenger fields are very light.  In this low cut-off regime we cannot use the effective terms to accurately calculate the octet production cross section \cite{Busoni:2014sya}.  However, we are assuming that if the octet prefers to decay via loop level couplings, it must have no tree level decays to messenger fields. That is, the mass of the octet must be less than twice the messenger mass $m_S<2M_{mess}$ and thus the messengers must be offshell in the process.  Gladly, we note that whenever this offshell condition,  $\sqrt{s}<2M_{mess}$,  is satisfied, that is exactly the region where the EFT is valid as well.  Since the effective operator coefficients are about equal to the UV messenger masses, may thus place a viability condition on the effective operator treatment that the coefficients should satisfy $ m_S<2\Lambda$. In Fig. \ref{cutvmass} we have drawn the line of effective operator viability, operator coefficients too light compared to the octet mass are inviable.

\section{Projection for LHC 14}

We will explore the reach of the 14 TeV run of LHC into our parameter space. Below we produce a sensitivity study for the process $gg \rightarrow S \rightarrow g\gamma$ in pp collisions at  $\sqrt{s}=14$ TeV. This search is based on the ATLAS 8 TeV $j \gamma$ resonance search.  In this analysis will follow the same cuts as the 8 TeV ATLAS analysis above,

\begin{itemize}
 \item At least one isolated photon with $p_{T}>125$ GeV
 \item At least one hard jet with $ p_{T}> 125$ GeV.
 \item Photon required to have angular separation $\Delta R(\gamma , j) > 1.0$ between leading photon and all other jets with $p_{T}>30$ GeV.
 \item Photon and jet required to be in the central region of the detector with $|\eta_{\gamma}|< 1.37$ and $|\eta_{j}|< 2.8$.
 \item Pseudo-rapidity separation between jet and photon of $|\Delta \eta (\gamma , j)| < 1.6 $.
 \item Highest $p_{T}$ $\gamma$ and jet candidates used to compute $m_{\gamma j}$, which is binned.
\end{itemize}

Our model is implemented in Feynrules.  For each point in our parameter space we generate signal events for our process using Madgraph5.  We shower the events  with Pythia and run them through the detector simulator PGS.  We then run a cut based the analysis on the events to determine the search efficiency.  In Figure 7a we show a plot of signal efficiency vs octet mass $m_S$. In general we find efficiencies similar to those of the 8 TeV analysis.  We may then calculate the total number of events $N_{evt}= \sigma \times L\times\epsilon \times BR_{g\gamma}$  in each invariant mass bin.

\begin{figure}[tbh-]
\centering
\includegraphics[scale=0.55]{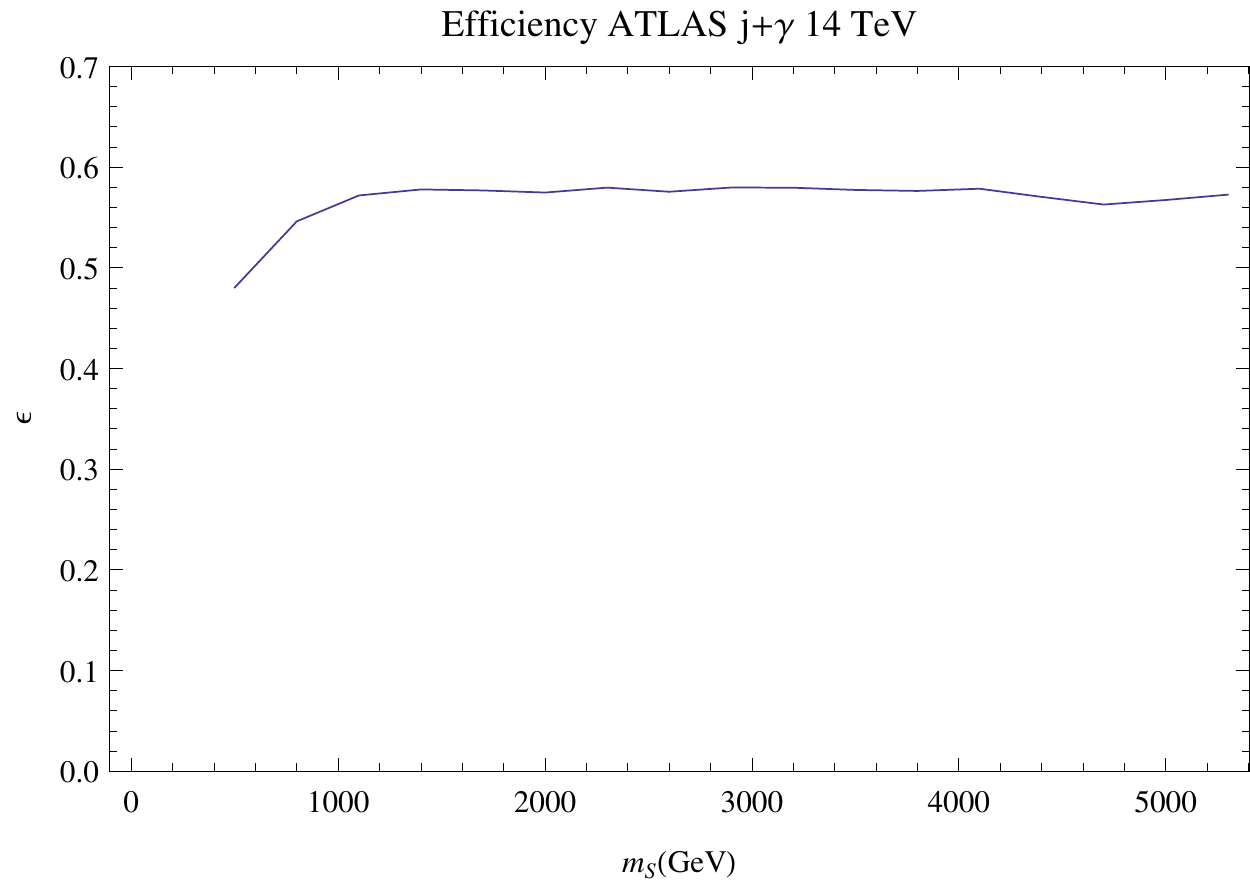}
\includegraphics[scale=0.89]{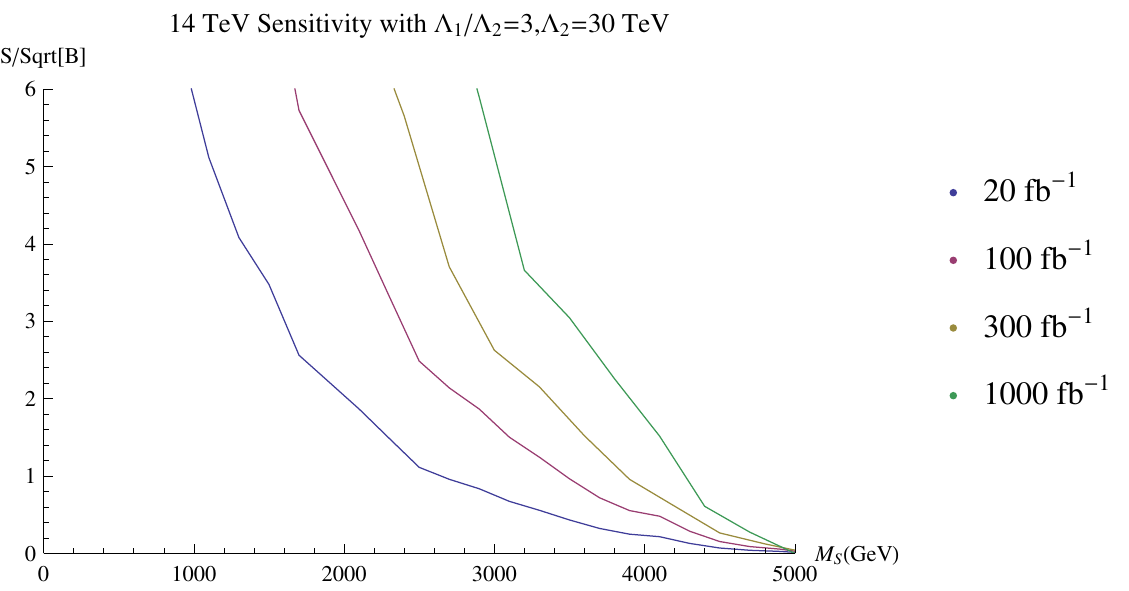}
\caption{
a.)On the left is the efficiency of signal events for $S\rightarrow g\gamma$ which pass the  selection criteria at $14$ TeV center of mass energy.
b.)On the right is the projected sensitivity for the 14 TeV run in the photon plus jet resonance channel .}
\label{fig:sensitivity}
\end{figure}

As in the 8 TeV search, the SM background for this process is dominated by $\gamma + j$ events. In order to estimate the background at 14 TeV we scale up the four parameter fit used to model the SM background in ATLAS's 8 TeV analysis.

We use a simple criteria for estimating sensitivity to new physics.  The signal to root background ratio  is denoted $S/\sqrt{B}$ where $S$ and $B$ are the expected number of signal and background events respectively.  For Poisson statistics, a search is sensitive to the existence of a new physics if $S/\sqrt{B} \geq 2$ and has discovery potential if $S/\sqrt{B} \geq 5$.

We plot the sensitivity for the 14 TeV process $gg \rightarrow S \rightarrow g\gamma$ in Fig \ref{fig:sensitivity}b. In the plot we fixed the benchmark effective operator coefficients $\Lambda_2=30$ TeV and $\Lambda_1=90$ TeV, and then plotted the expected $S/\sqrt{B}$ ratio vs. the octet mass $m_S$.  We have shown the sensitivity for various values of the run luminosity.  For this benckmark point, we find that a octets of $m_{S}\lesssim 1.8$ TeV have 5 sigma discovery potential in $100$ fb$^{-1}$ of luminosity.  In $1$ ab$^{-1}$  octets of  $m_{S}\lesssim 3$ TeV have discovery potential in this channel.  In the appendix we shown the reach in the $\Lambda_1-\Lambda_2$ plane for various values of the octet mass.

\vspace{5mm}
\section{Conclusions}

We have considered a set of effective dimension 5 operators through which standard model adjoint scalars couple to pairs of SM gauge bosons. These operators may present themselves in a variety of models. We have considered supersymmetric and non-supersymmetric UV completions for these scenarios including the popular set of 'R-symmetric' SUSY models, which demand the existence of scalar fields which are adjoints under the SM gauge groups.  These operators open up new phenomenological possibilities in the search for scalar adjoint states.

In this work we study the phenomenology of the production and decay of a single color octet scalar which couples to pairs of SM gauge bosons, $gg$, $Z \gamma$, $Z g$.  We have calculated bounds on this scenario from several analyses from the 8 TeV run of LHC. We have placed lower limits on effective operator coefficients for a variety of octet masses using $\gamma$+jet, dijet, monojet and V+j search channels. We have also explored the discovery potential of the octet scalar in the gluon-photon final state for the 14 TeV run of LHC.

These operators open the possibility for a variety of new production and decay channels for scalar octet searches.
For example, the decays of the octet scalar in the Z g channel alone include such final state topologies as jet plus a hard dilepton pair, as well as jets plus $\met$, and jet+hadronic tagged Z topologies.  Though the branching fraction into leptons is small, this may be an interesting low background channel to search in for new states. Similarly interesting decay channels exist for equivalent operators pertaining to SU(2) adjoint scalars. We also note that we have here considered only single scalar octet production through gluon fusion.  There are, however, stranger production channels to  be studied.
For example, even if the Sgg coupling were somehow small, the $Sg\gamma$ coupling would allow S production through the processes $gg>g^{*}>S \gamma$ and even  $gg>g^{*}>S  Z$.  Models with such adjoints have a variety of collider topologies which would be an interesting topic of further study.

As discussed in the section on high energy models, the dimension 7 operators listed in Eqn 3. are produced at loop level by integrating out heavy squarks/sleptons in R-symmetric SUSY models.  This set of operators presents more possibilities for phenomenological study if the operators are not too suppressed.  For example once Higgs vevs are inserted, the operators allow production of the SU(2) adjoint though gluon fusion. The decay of this adjoint may have a striking leptonic signature.  More complex scenarios which involve SM adjoints may have even more striking signatures.  For example, in the Monohar-Wise model there exists scalar color octets which are fundamentals and anti-fundamentals of SU(2). Effective couplings of these adjoints to pairs of vector bosons may lead to very interesting decay chains.

Finally we note that UV completions of our models predict correlations of signals across multiple channels.  A combinations of final state searches will be the most powerful tool to rule on the parameter space of these models.  This will also help to determine to the correct UV physics if a scalar adjoint signal is seen in a channel.  For example, in an R-symmetric-MSSM scenario the detection of a scalar color octet in the dijet resonance channel would demand a specific rate in the j+photon channel as well.  Measuring these rate would give valuable information about the messenger and squark sectors of the theory.

\section{Appendix}

For the Monohar Wise model, a color octet Scalar is also an SU(2) doublet $S_u= (8,2)_1, S_d= (8,\overline{2})_{-1}$ \cite{Manohar:2006ga}.  We may write the most general terms effective Lagrangian which couples these fields to pairs of SM gauge bosons,

\begin{equation}
\begin{split}
L&= \frac{1}{\Lambda_{gg1}^2}d^{abc} H^{\dagger} S_{ua} G_{b}^{\mu \nu}G_{c}^{\mu \nu}+ \frac{1}{\Lambda_{gg2}^2} d^{abc}H S_{da} G_{b}^{\mu \nu}G_{c}^{\mu \nu} + \frac{1}{\Lambda_{gb1}^2} H^{\dagger} S^a_u G_{a}^{\mu \nu}B^{\mu \nu}+ \frac{1}{\Lambda_{gb2}^2} HS_d G_{a}^{\mu \nu}B^{\mu \nu} \\ \nonumber
&+ \frac{1}{\Lambda_{gw1}^2} [H^{\dagger}W^{\mu \nu}S_u] G_{a}^{\mu \nu}+ \frac{1}{\Lambda_{gw2}^2} [H W^{\mu \nu}S_d] G_{a}^{\mu \nu}
\end{split}
\end{equation}
Were $G^{\mu \nu}$ is the SU(3) field strength tensor,  $W^{\mu \nu}$ is the SU(2) field strength tensor and  $B^{\mu \nu}$ is the U(1) fields strength tensor.  In the first 4 terms SU(2) indices of the scalar octets are contracted with those of the Higgs.  In the last two operators SU(2) indices are contracted between the Higgs, scalar octet and $W^{\mu \nu}$ as indicated by brackets.

Finally we may consider the effective Lagrangian of a bi-adjoint under SU(3) and SU(2)

\begin{equation}
L= \frac{1}{\Lambda_{sgw}} S_{a}^{j} G_{a}^{\mu \nu} W_{\mu \nu}^{i} + \frac{1}{\Lambda_{sgb}^3} [H^{\dagger} S_{a}^{j}H] G_{a}^{\mu \nu} B_{\mu \nu}^{i} + \frac{1}{\Lambda_{sgg}^3}  [H^{\dagger} S_{a}^{j}H] G_{a}^{\mu \nu} G_{\mu \nu}^{i}
\end{equation}

Below we give exact branching fractions for a scalar color octet into gg, $g\gamma$ and gZ final states.  Couplings between the octet and pairs of gauge bosons follow from Eqn 1.
\begin{equation}
\begin{split}
&BR_{g\gamma}=
\frac{c_{W}^{2}}
{c_{W}^{2}
+s_{W}^{2}(1-m_{Z}^{2}/m_{S}^{2})^{3}
+\frac{10}{3}\Lambda_{1}^{2}/\Lambda_{2}^{2}} \\
&BR_{gZ}=
\frac{s_{W}^{2}(1-m_{Z}^{2}/m_{S}^{2})^{3}}
{c_{W}^{2}
+s_{W}^{2}(1-m_{Z}^{2}/m_{S}^{2})^{3}
+\frac{10}{3}\Lambda_{1}^{2}/\Lambda_{2}^{2}} \\
&BR_{gg}=
\frac{1}
{1+
\frac{3}{10}\Lambda_{2}^{2}/\Lambda_{1}^{2}c_{W}^{2}
+\frac{3}{10}\Lambda_{2}^{2}/\Lambda_{1}^{2}s_{W}^{2}(1-m_{Z}^{2}/m_{S}^{2})^{3}},
\end{split}
\end{equation}

Where S is the octet, the $\Lambda_i$'s are the effective operator cutoffs, and $c_{W}$ and $s_{W}$  are the cosine and sine of the weak angle.

\begin{figure}[tbh-]
\centering
\includegraphics[scale=0.95]{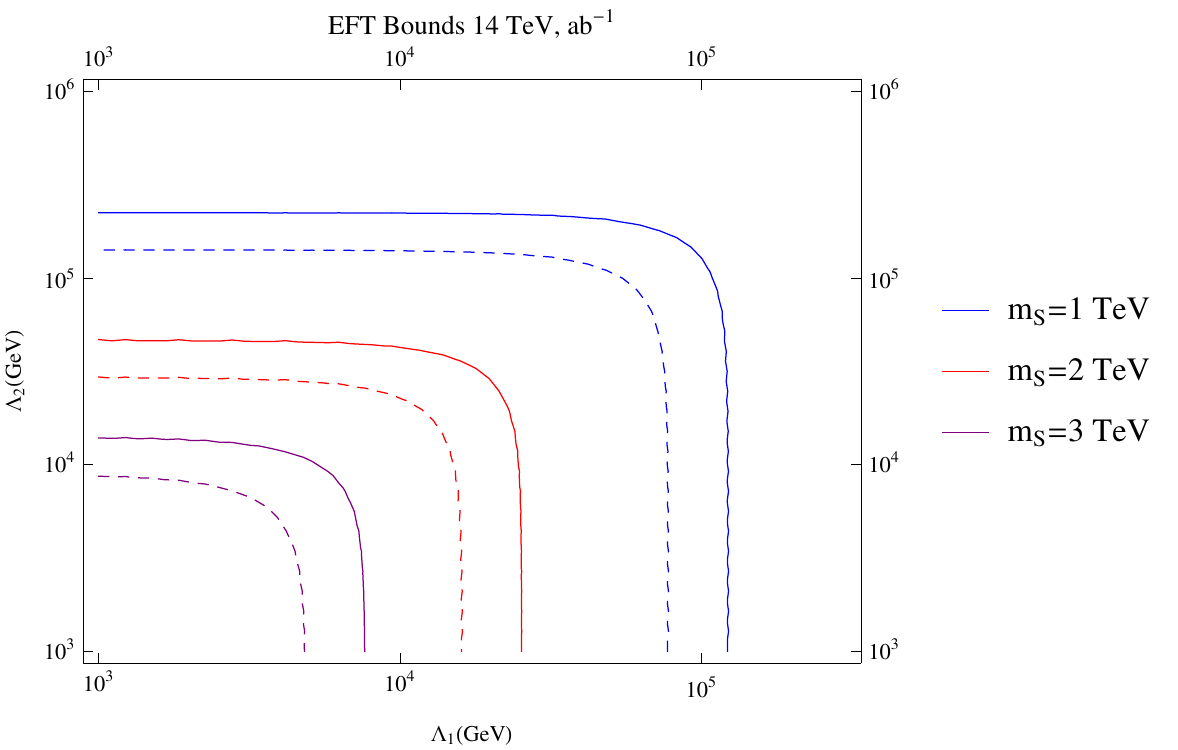}
\caption{
Projected sensitivity of color octet searches in the $g\gamma$ channel for the 14 TeV LHC run. Shown are reached for various valued of the octet mass in the $\Lambda_2-\Lambda_1$ plane.  Solid lines indicate the 2 sigma reach while dotted lines indicate 5 sigma reach.}
\label{fig:sdfsdf}
\end{figure}

\section{Acknowledgements}
This work was made possible by fund from DOE grant DE-SC001352.


\end{document}